\documentclass[prb,amsmath,showpacs,draft,amsmath,amsfonts,twocolumn,superscriptaddress]{revtex4}

\usepackage{bm}
\begin{document}

\title{Unified hydrodynamics theory of the lowest Landau level}

\author{I.~V.~Tokatly}

\email{ilya.tokatly@physik.uni-erlangen.de}

\affiliation{Lerhrstuhl f\"ur Theoretische Festk\"orperphysik,
  Universit\"at Erlangen-N\"urnberg, Staudtstrasse 7/B2, 91058
  Erlangen, Germany}

\affiliation{Moscow Institute of Electronic Technology,
 Zelenograd, 124498 Russia}
\date{\today}

\begin{abstract}
We propose a hydrodynamics theory of collective quantum Hall states,
which describes incompressible liquids, hexatic liquid crystals, a
bubble solid and a Wigner crystal states within a unified
framework. The structure of the theory is uniquely determined by the
space-time symmetry, and a symmetry with respect to static shear
deformations. In agreement with recent experiments the theory predicts
two gapped collective modes for incompressible liquids. We argue that
the presence of the above two modes is a universal property of a
magnetized two-dimensional collective liquid.
\end{abstract}


\maketitle

\section{Introduction}

Hydrodynamics or, more generally, continuum mechanics is a common tool
to describe collective degrees of freedom in condensed matter
systems. This approach becomes especially powerful in the linear
regime as it provides a unified view on collective dynamics in solids,
liquid crystals, normal liquids, superfluids, etc. (see, for example,
Ref~\onlinecite{MarParPer1972}). The hydrodynamics equations of motion follow
the universal conservation laws, while a particular state of matter
enters the theory via a linearized ``equation of state'' that is
parametrized by a set of visco-elastic constants. A general form of
the ``equation of state'' (normally this is a dependence of the stress
tensor on basic variables), and a number of independent parameters are
determined by the macroscopic symmetry. 

Hydrodynamics represents the response of a many-particle system to
external fields in terms of a few collective variables. This assumes
that for whatever reason the underlying particles loose their
individual properties and get collectivized. Apparently the above
assumption is perfectly valid for quantum collective many-body states,
such as superfluids or strongly correlated states of an electron gas
in the fractional quantum Hall (FQH) regime
\cite{QHEbyChakraborty,QHEbyYoshioka}. The hydrodynamics of
superfluids has indeed a long and reach history (a general review as
well as recent applications to trapped atomic Bose condensates can be
found in Refs.~\onlinecite{PethickSmith,BEC2003}). Yet the power of
hydrodynamics approach in application to FQH fluids has been not fully
explored. A notable exception is a low energy (in fact, adiabatic)
extreme, where the structure of hydrodynamic theory is well
understood. In this regime the hydrodynamics of FQH liquids
universally reduces to a topological Chern-Simons theory
\cite{Wen1995} (see also Ref.~\onlinecite{Susskind2001}), which
reflects a topological order of incompressible FQH states. An
effective Chern-Simons theory successfully describes gapless edge
excitations, but, by its low-energy nature, fails to reproduce
fundamental bulk modes with a gapped spectrum \cite{GirMacPla1986}. 

A principal possibility to describe gapped magneto-roton modes within
a continuum mechanics formalism has been demonstrated recently
\cite{TokPRB2006}. The most important observation of this work is that
the gap $\Delta$ in the spectrum of collective excitations requires a
highly unusual, resonant frequency dependence of the shear stress. An
effective dynamic shear modulus should diverge at $\omega=\Delta$,
otherwise collective modes will be gapless. It has been shown that the
required resonant structure of the shear modulus naturally appears in
hydrodynamics with an additional tensor collective variable that
describes nontrivial precession dynamics of a shear stress. Using
heuristic physical arguments in Ref.~\onlinecite{TokPRB2006} we
constructed the simplest equation of motion for the new tensor
variable, and thus derived a ``minimal'' hydrodynamic theory that
reproduces a qualitatively correct dispersion of a gapped collective
mode.

The present paper reports a further progress in that
direction. We show that there is a fundamental local symmetry behind
the presence of a tensor field in the hydrodynamics of incompressible
FQH liquids. That is an invariance with respect to arbitrary static
shear deformations, which is a universal property of a liquid state of
matter. The tensor dynamic variable is, in fact, a gauge field that
supports the above local symmetry. 

From experimental side an interest to bulk collective excitations of
FQH liquids has been recently renewed by the first observation of the
long wavelength dispersion of gapped modes at the filling factor
$\nu=1/3$.  \cite{Hirjibehedin2005} The most intriguing result of
these experiments is a two-branch structure of the gapped excitation
spectrum at small wave vectors. 

In the present paper we propose a
universal hydrodynamical interpretation of these experimental
findings. We derive a hydrodynamics of FQH liquids, which relies
solely on general symmetry arguments. The only assumption, which
underlies our theory, is that the system is a two-dimensional (2D)
collective liquid with a proper space-time symmetry
(i.~e. rotationally and time-reversal invariant).
Incompressibility of the liquid, nonvanishing dissipationless Hall
conductivity, a gapped spectrum of collective modes, etc., follow
automatically from that assumption.  The theory predicts two gapped
modes (one with upward, and another one with downward dispersions),
which is in excellent agreement with recent experiments
\cite{Hirjibehedin2005}. The key idea in constructing the
hydrodynamics of incompressible liquids is to preserve the fundamental
property of any liquid state -- a local symmetry under static
volume-preserving deformations. By a successive reduction of the above
local invariance we deduce theories of a hexatic liquid crystal, and
two different crystalline phases.

The paper is organized as follows. In Sec.~II we identify basic
symmetries, and present a phenomenological derivation of the
linearized hydrodynamics of incompressible QH liquids. We construct a
Lagrangian of the theory, derive equations of motion for basic
collective variables, and analyze the spectrum of collective
excitations. In Sec.~III the theory is generalized to describe liquid
crystal, and crystal states with hexagonal symmetry. In Sec.~IV we
summarize our main results.

\section{Hydrodynamics of incompressible quantum Hall liquids} 

\subsection{Lagrangian of a magnetized collective liquid}

Our goal is to describe linear dynamics of an interacting 2D
many-particle system confined to the $x,y$-plane, and subjected to a
strong transverse magnetic field ${\bf B}={\bf e}_{z}B$. The
hydrodynamic theory, which we derive in this section, is based on the
following two, quite general assumptions about the ground/equilibrium
state of the system: (i) The system is in a collective state that is
rotationally, time-reversal \cite{note1}, and Galilee invariant; (ii)
The system is in a liquid state.

A few comments on the above assumptions are in order. The term
collective state assumes the absence of a continuum of single particle
excitations. Hence the dynamics can be represented by a finite number of
collective variables. One of those variables, namely the displacement
vector ${\bf u}({\bf r},t)$, is fixed as it should appear in any
linearized continuum mechanics. The velocity field ${\bf v}({\bf
r},t)$, and the variation of the density $\delta n({\bf r},t)$ are
related to the displacement ${\bf u}$ as follows, 
$$
{\bf v}=\partial_{t}{\bf u}, \qquad
\delta n =-n\nabla{\bf u},
$$ 
where $n=\nu/2\pi l^{2}$ is the equilibrium density ($\nu$ is the
filling factor, and $l$ is the magnetic length). In general the
displacement does not exhaust all relevant dynamic variables. It is
well known that an adequate hydrodynamic description of ordered
states of matter requires an introduction of additional fields that
are specific for every system (a magnetization in ferromagnets, a
director vector in nematic liquid crystals, a superfluid velocity in
superfluids, etc.). We will see that a somewhat similar (though not
precisely the same) situation takes place for incompressible FQH
liquids.

The assumption of a liquid state, (ii), can be also formalized in
terms of the displacement vector. By definition, a liquid is a state
of matter that does not respond to an arbitrary static shear
(i.~e. volume-preserving) deformation. Mathematically shear
deformations are described a divergenceless displacement vector. Hence
by displacing a liquid system with a static, purely transverse vector
${\bf f}({\bf r})$ we should not produce any physical effect. Formally
this means that the equations of motion for any 2D liquid should be
invariant under the following transformation \cite{note3}
\begin{equation}
u_{i}'({\bf r},t) = u_{i}({\bf r},t) + 
\varepsilon_{ij}\partial_{j}\psi({\bf r}), 
\label{1}
\end{equation}
where $\psi({\bf r})$ is an arbitrary function of only spatial
variables. The invariance under a volume preserving diffeomorphism of
Eq.~(\ref{1}) has a transparent interpretation within a Lagrangian
formulation of fluid mechanics. It reflects the symmetry with respect
to relabeling of different infinitesimal fluid elements
\cite{Jackiw2004}. It is worth noting that the above diffeomorphism
symmetry can serve as a most general mathematical definition of a
liquid state. In terms of visco-elastic coefficients the symmetry
under the transformation of Eq.~(\ref{1}) implies a vanishing static
shear modulus.

An important requirement, which we impose on the theory, follows from
the universality of the high-frequency response. In the
high-frequency, long wavelength limit any interacting many-body system
behaves as an elastic continuum (see, for example,
Ref.~\onlinecite{GiulianiVignale}, and a recent paper of
Ref.~\onlinecite{TokPRB2005b} for a most general demonstration of this
fact). Formally this means that in the high-frequency regime the exact
stress tensor takes the standard elastic form \cite{LandauVII:e}
$$
P_{ij} = -\delta_{ij}K\nabla{\bf u} - \mu_{\infty}
(\partial_{i}u_{j}+\partial_{j}u_{i}-\delta_{ij}\nabla{\bf u}), 
$$
where the high-frequency shear modulus
$\mu_{\infty}$, and the bulk modulus $K$ are universal constants that
are related to the ground state properties \cite{TokPRB2005b}. 
Therefore in the limit
$\omega\to\infty$ the theory should reduce to the following equation
for displacement vector ${\bf u}$, which describes linear dynamics of
the classical elastic medium \cite{LandauVII:e}
\begin{equation}
m\partial_{t}^{2}{\bf u} +B\partial_{t}{\bf u}\times{\bf e}_{z} -
\frac{K}{n}\nabla(\nabla{\bf u}) -
\frac{\mu_{\infty}}{n}\nabla^{2}{\bf u} + \nabla U_{\rm H} =
{\bf F},
\label{2}
\end{equation}
where $U_{\rm H}$, and ${\bf F}$ are the Hartree potential, and the
external force respectively. The
high-frequency equation of motion, Eq.~(\ref{2}), is valid for all
rotationally and Galilee invariant systems in the presence of an
external magnetic field. 
The above requirement uniquely determines the high-frequency form of the
Lagrangian that generates the correct left hand side of the equation of motion,
Eq.~(\ref{2})  
\begin{equation}
L^{\infty} = L_{0} + L^{\infty}_{\rm S},
\label{3}
\end{equation}
where $L_{0}$ is a linearized Lagrangian of an ideal liquid \cite{note2}
\begin{equation}
L_{0} = \frac{1}{2}nm(\partial_{t}{\bf u})^{2} - 
       \frac{1}{2}nB{\bf u}(\partial_{t}{\bf u}\times{\bf e}_{z}) -
      \frac{1}{2}K(\nabla{\bf u})^{2},
\label{4}
\end{equation}
and $L^{\infty}_{\rm S}$ is the standard elastic shear Lagrangian 
\begin{equation}
L^{\infty}_{\rm S} = -\frac{1}{2}\mu_{\infty}(\partial_{i}u_{j})^{2}.
\label{5a}
\end{equation}
For the further discussion it convenient to introduce a traceless
shear strain tensor
\begin{equation}
S_{ij}=\partial_{i}u_{j}+\partial_{j}u_{i}-\delta_{ij}\nabla{\bf u},
\label{5b}
\end{equation}
and to represent the shear Lagrangian of Eq.~(\ref{5a}) in the following
``natural'' form
\begin{equation}
L^{\infty}_{\rm S} = -\frac{1}{4}\mu_{\infty}{\rm Tr}\hat{\bf S}^{2}\equiv
-\frac{1}{4}\mu_{\infty} S_{ij}S_{ji},
\label{5}
\end{equation}
One can straightforwardly check that Eqs.~(\ref{5a}) and (\ref{5}) are
identical up to an irrelevant total derivative.  It is important to
outline that the high-frequency form of the Lagrangian, Eq.~(\ref{3}),
is not an additional assumption, but the exact property of any
rotationally and Galilee invariant interacting many-body system.

Let us analyze the symmetries of the Lagrangian $L^{\infty}$ defined
by Eqs.~(\ref{3}), (\ref{4}), and (\ref{5}). Obviously it is
rotationally, time-reversal, and Galilee invariant. [It is worth
mentioning that the requirement of Galilee invariance fixes the
coefficients in the first two terms in Eq.~(\ref{4}) to $nm$ and $nB$
respectively.] Hence the our first assumption, (i), is fulfilled at
the high frequency level. However, the assumption (ii), which
corresponds to the local invariance of Eq.~(\ref{1}), is violated by
the presence of the shear term in $L^{\infty}$. Apparently the first
term, $L_{0}$, in Eq.~(\ref{3}) is invariant under the transformation
Eq.~(\ref{1}). The second, shear term, $L^{\infty}_{\rm S}$, fails to
satisfy this property since the shear strain tensor, Eq.~(\ref{5b}),
acquires a correction
\begin{eqnarray}
\hat{\bf S}'({\bf r},t) &=&
 \hat{\bf S}({\bf r},t) + \delta\hat{\bf S}({\bf r})
\label{6a}\\
\delta S_{ij}({\bf r}) &=&  
\varepsilon_{ik}\partial_{k}\partial_{j}\psi({\bf r}) + 
\varepsilon_{jk}\partial_{k}\partial_{i}\psi({\bf r}),  
\label{6}
\end{eqnarray}
which changes $L^{\infty}_{\rm S}$. A common prescription to restore any
local symmetry is offered by the gauge theory -- one has to introduce
a gauge field $\hat{\bm\eta}({\bf r},t)$ that compensates unwanted
changes in the Lagrangian. Namely, we replace $\hat{\bf S}^{2}$ in
Eq.~(\ref{5}) by the form $(\hat{\bf S} - \hat{\bm\eta})^{2}$, and add
a Lagrangian $L_{\bm\eta}$ that describes dynamics the gauge field
$\hat{\bm\eta}$, i.~e.
$$
L^{\infty}_{\rm S} \to L^{\rm liq}_{\rm S} =
-\frac{1}{4}\mu_{1}{\rm Tr}(\hat{\bf S} - \hat{\bm\eta})^{2}
+ L_{\bm\eta}.
$$
The local symmetry of Eq.~(\ref{1}) is guaranteed if the term
$L_{\bm\eta}$ in the Lagrangian is invariant under the transformation
\begin{equation}
\hat{\bm\eta}'({\bf r},t)= \hat{\bm\eta}({\bf r},t) 
+ \delta\hat{\bf S}({\bf r}),
\label{6b}
\end{equation}
where $\delta\hat{\bf S}({\bf r})$ is defined after
Eq.~(\ref{6}). Additional restrictions to the form of $L^{\rm
liq}_{\rm S}$ are imposed by the space-time symmetry, and a fixed
high-frequency asymptotic form of Eq.~(\ref{5}). Apparently, since the
shear strain tensor $\hat{\bf S}$, Eq.~(\ref{5b}), is symmetric and
traceless, the gauge field $\hat{\bm\eta}$ also has to be a symmetric,
traceless tensor. Following the outlined route we construct the
following most general (quadratic) shear Lagrangian that is
rotationally and time-reversal invariant, and enjoys the local
symmetry of Eq.~(\ref{1}):
\begin{eqnarray} \nonumber
L^{\rm liq}_{\rm S} =       
&-& \frac{1}{4}\mu_{1}{\rm Tr}(\hat{\bf S} - \hat{\bm\eta})^{2}
 + \frac{1}{4}\frac{\mu_{1}}{\Delta_{1}^{2}}\Big[
   {\rm Tr}(\partial_{t}\hat{\bm\eta})^{2}\\
&+& \Delta_{2}{\rm Tr}\hat{\bm\eta}(\partial_{t}\hat{\bm\eta}\times{\bf e}_{z}) 
 + 2\Delta'{\rm Tr}\hat{\bf S}(\partial_{t}\hat{\bm\eta}\times{\bf e}_{z})
\Big],
\label{7}
\end{eqnarray}
where the cross product of ${\bf e}_{z}$ and a symmetric tensor 
is defined as follows 
$$
(\hat{\bm\eta}\times{\bf e}_{z})_{ij} = 
(\varepsilon_{ik}\eta_{kj} + \varepsilon_{jk}\eta_{ki})/2.
$$
The term in the square brackets in Eq.~(\ref{7}) contains all allowed
scalar invariants up to the second order in space-time derivatives. It
should be noted that the rotational and the time-reversal symmetries
allow two more quadratic invariants:
\begin{equation}
 {\rm Tr}\hat{\bm\eta}^{2}, \quad
{\rm Tr}(\partial_{i}\hat{\bm\eta})(\partial_{i}\hat{\bm\eta}).
\label{7a}
\end{equation}
However these terms are forbidden by the invariance under the local
transformation Eq.~(\ref{6b}). As a result only the terms with at
least one time derivative are left in $L_{\bm\eta}$.

The final Lagrangian of a collective quantum Hall liquid takes the form
\begin{equation}
L^{\rm liq} = L_{0} + L^{\rm liq}_{\rm S} 
+ n{\bf u}\partial_{t}{\bf a} - n{\bf u}\nabla\varphi
\label{8}
\end{equation}
where $L_{0}$ and $L^{\rm liq}_{\rm S}$ are given by Eqs.~(\ref{4})
and (\ref{7}) respectively, and the last two terms describe the
interaction with external potentials, ${\bf a}({\bf r},t)$ and
$\varphi({\bf r},t)$. 

The Lagrangian Eq.~(\ref{8}) defines a theory of two coupled fields --
the vector field ${\bf u}({\bf r},t)$ and a symmetric traceless tensor
field $\hat{\bm\eta}({\bf r},t)$. Both our basic assumptions, (i) and
(ii), are satisfied by Eq.~(\ref{8}). This is the most general
harmonic Lagrangian for a 2D collective liquid in the presence of a
transverse magnetic field. The theory contains four still
undefined phenomenological parameters, $\mu_{1}$, $\Delta_{1}^{2}$,
$\Delta_{2}$ and $\Delta'$. The requirement of the high-frequency
elastic response provides one constraint, which allows us to express
$\mu_{1}$ in terms of $\mu_{\infty}$. Three remaining parameters can
be related to observable quantities -- the excitation gaps, and the
corresponding oscillator strength (see Sec.~IIB below).

\subsection{Equations of motion and response functions}

The Lagrangian Eq.~(\ref{8}) generates the following equation of motion
for the displacement vector
\begin{equation}
mn\partial_{t}^{2}{\bf u} + nB\partial_{t}{\bf u}\times{\bf e}_{z} -
 K\nabla(\nabla{\bf u}) + \nabla\hat{\bm\pi} = n{\bf F},  
\label{9}
\end{equation}
where ${\bf F}=\partial_{t}{\bf a}-\nabla\varphi$, and $\hat{\bm\pi}$
is the traceless shear stress tensor that is defined as follows
\begin{equation}
\hat{\bm\pi} = -\mu_{1}\big[\hat{\bf S} - \hat{\bm\eta}
 - (\Delta'/\Delta_{1}^{2})\partial_{t}\hat{\bm\eta}\times{\bf
   e}_{z}
\big]  
\label{10}
\end{equation}
Similarly, variation of the Lagrangian with respect to
$\hat{\bm\eta}$ yields the equation of motion for the gauge field
\begin{equation}
\partial_{t}^{2}\hat{\bm\eta} +\Delta_{1}^{2}\hat{\bm\eta} 
- \Delta_{2}\partial_{t}\hat{\bm\eta}\times{\bf e}_{z} = 
\Delta_{1}^{2}\hat{\bf S} + \Delta'\partial_{t}\hat{\bf S}\times{\bf e}_{z}.
\label{11}
\end{equation}
The tensor field $\hat{\bm\eta}$ enters the equation of motion for the
displacement vector ${\bf u}$ only via the stress tensor $\hat{\bm\pi}$,
Eq.~(\ref{10}). Therefore it is more convenient technically to
consider $\hat{\bm\pi}$ as an independent dynamic variable.
Making use of Eqs.~(\ref{10}) and (\ref{11}) we can eliminate the
gauge field $\hat{\bm\eta}$ in favor of $\hat{\bm\pi}$, and thus derive
the following equation of motion for the shear stress tensor 
\begin{eqnarray}\nonumber
\partial_{t}^{2}\hat{\bm\pi} + \Delta_{1}^{2}\hat{\bm\pi} 
&-& \Delta_{2}\partial_{t}\hat{\bm\pi}\times{\bf e}_{z} = 
-\mu_{1}\left(1 + \Delta'^{2}/\Delta_{1}^{2}\right)
\partial_{t}^{2}\hat{\bf S}\\ &-&
\mu_{1}(\Delta_{2} + 2\Delta')\partial_{t}\hat{\bf S}\times{\bf e}_{z}.
\label{12}
\end{eqnarray}
Using Eq.~(\ref{12}) we can identify one of the phenomenological
parameters entering the theory, namely the parameter $\mu_{1}$.  In
the high-frequency regime the equation of motion for the displacement,
Eq.~(\ref{9}), should reduce to its asymptotic form,
Eq.~(\ref{2}). This means that at $\omega\to\infty$ the solution to
Eq.~(\ref{12}) should take the form
$\hat{\bm\pi}=-\mu_{\infty}\hat{\bf S}$. In the $\omega\to\infty$
regime the behavior of Eq.~(\ref{12}) is dominated by terms with the
highest order of time derivatives, i.~e. by the first terms in the
right and left hand sides of Eq.~(\ref{12}). Thus the requirement of
the elastic high-frequency response fixes parameter $\mu_{1}$ in the
Lagrangian to the value
\begin{equation}
\mu_{1}  = \mu_{\infty}(1 + \Delta'^{2}/\Delta_{1}^{2})^{-1}.
\label{13}
\end{equation}

The system of Eqs.~(\ref{9}), (\ref{12}) completely determine the
dynamics of the system. Using Eqs.~(\ref{9}) and (\ref{12}) we can calculate
the response to any configuration of external fields, and find all
collective modes of the system. Let us first analyze the equation of
motion for $\hat{\bm\pi}$, Eq.~(\ref{12}). This equation
determines a dynamic ``equation of state'' that relates the stress
tensor $\hat{\bm\pi}$ to the strain tensor $\hat{\bf S}$.
By symmetry the Fourier component of the shear stress tensor $\hat{\bm\pi}$
should be representable in the form \cite{TokPRB2006}
\begin{equation}
\hat{\bm\pi} = -\mu(\omega)\hat{\bf S} + 
i\omega\Lambda(\omega)\hat{\bf S}\times{\bf e}_{z},
\label{14}
\end{equation}
where $\mu(\omega)$ is the dynamic shear modulus, and
$\Lambda(\omega)$ is a ``magnetic'' modulus that is responsible for a
Lorentz shear stress. Substituting the representation of
Eq.~(\ref{14}) into Eq.~(\ref{12}) we arrive at the following system of
two equations for functions $\mu(\omega)$ and $\Lambda(\omega)$
\begin{eqnarray}
(\omega^{2} - \Delta_{1}^{2})\mu &+& \omega^{2}\Delta_{2}\Lambda 
 = \omega^{2}\mu_{\infty},
\label{12a}\\
(\omega^{2} - \Delta_{1}^{2})\Lambda &+& \Delta_{2}\mu = -\mu_{\infty}
\frac{\Delta_{2} + 2\Delta'}{1 + \Delta'^{2}/\Delta_{1}^{2}},
\label{12b}
\end{eqnarray}
where we have used the relation of $\mu_{1}$ to $\mu_{\infty}$,
Eq.~(\ref{13}). Straightforward solution of Eqs.~(\ref{12a}),
(\ref{12b}) yields the dynamic moduli $\mu(\omega)$ and
$\Lambda(\omega)$ for the theory defined by the Lagrangian
Eq.~(\ref{8})
\begin{eqnarray}
\mu(\omega) &=& \mu_{\infty}\omega^{2}\left[
\frac{f_{-}}{\omega^{2}-\Delta_{-}^{2}} + 
\frac{f_{+}}{\omega^{2}-\Delta_{+}^{2}}\right],
\label{15}\\
\Lambda(\omega) &=& \mu_{\infty}\left[
\frac{f_{-}\Delta_{-}}{\omega^{2}-\Delta_{-}^{2}} - 
\frac{f_{+}\Delta_{+}}{\omega^{2}-\Delta_{+}^{2}}\right],
\label{16}  
\end{eqnarray}
The quantities $\Delta_{+}$ and $\Delta_{-}$ are related to
parameters $\Delta_{1}^{2}$ and $\Delta_{2}$ in the shear Lagrangian of
Eq.~(\ref{7}) as follows 
\begin{equation}
\Delta_{+}\Delta_{-} = \Delta_{1}^{2}, \qquad 
\Delta_{+} - \Delta_{-} = \Delta_{2}
\label{17}
\end{equation}
The oscillator strengths $f_{\pm}$ are given by the
expressions 
\begin{equation}
f_{\pm} = \frac{\Delta_{\pm}(\Delta_{\mp} \mp \Delta')^{2}}{(\Delta_{+}
 + \Delta_{-})(\Delta_{+}\Delta_{-} +\Delta'^{2})},
\label{18}
\end{equation}
and satisfy the sum rule $f_{+}+f_{-}=1$. 

The frequency dependence of shear modulus $\mu(\omega)$,
Eq.~(\ref{15}), agrees with our general expectations. It vanishes at
$\omega\to 0$, and approaches $\mu_{\infty}$ at $\omega\to\infty$. A
resonant structure of Eqs.~(\ref{15}) and (\ref{16}) is responsible
for a gapped spectrum of collective excitations \cite{TokPRB2006}
(we will see that $\Delta_{\pm}$ are, in fact, the excitation gaps).

To calculate the response functions we need to substitute the stress
tensor, Eq.~(\ref{14}), into the equation for ${\bf u}$,
Eq.~(\ref{9}), and solve it for given external potentials ${\bf a}$
and $\varphi$. Below we consider the most interesting limit of a strong
magnetic field when the first (acceleration) term in Eqs.~(\ref{4})
and (\ref{9}) becomes irrelevant. Physically this corresponds the
situation with all particles occupying the lowest Landau level
(LLL). Formally the intra-LLL dynamics is described by the Lagrangian
Eq.~(\ref{8}) [or, equivalently, by the equations of motions
Eqs.~(\ref{9}), (\ref{12})] in the limit of vanishing bare mass, $m\to
0$. Substituting Eq.~(\ref{14}) into Eq.~(\ref{9}), and neglecting the
acceleration term we get the equation that describes the
LLL-projected dynamics of the displacement vector ${\bf u}({\bf
  q},\omega)$ 
\begin{equation} 
-i\omega (nB + \Lambda q^{2}){\bf
u}\times{\bf e}_{z} + \mu q^{2}{\bf u} + K{\bf q}
({\bf qu}) = n{\bf F},
\label{LLL}
\end{equation} 
where $\mu$ and $\Lambda$ are given by Eqs.~(\ref{15}) and (\ref{16}),
and the force in the right hand side is related to the external potentials as
follows, ${\bf F}=-i\omega {\bf a} - i{\bf q}\varphi$.

Let us first calculate the Hall conductivity
$\sigma_{H}(\omega,q)$. By definition, $\sigma_{H}(\omega,q)$ relates
the $x$-component of the current to the $y$-component of an electric
field,
$$
j_{x}=\sigma_{H}(\omega,q)E_{y}
$$  
Noting that ${\bf j}= n{\bf v}= -i\omega n{\bf u}$ and solving
Eq.~(\ref{LLL}) with ${\bf a}=0$, and $-i{\bf q}\varphi={\bf E}$ we
find
\begin{equation}
\sigma_{H}(\omega,q) = n^{2}
\frac{\omega^{2}(nB + \Lambda q^{2})}{\omega^{2}(nB + \Lambda q^{2})^{2}
- \mu(\mu + K)q^{4}}.
\label{19}
\end{equation}
The dissipationless (equilibrium) Hall conductivity is calculated from
Eq.~(\ref{19}) by taking first the limit $\omega\to 0$, and then the limit
$q\to 0$. Since $\mu(\omega)$, Eq.~(\ref{15}), vanishes as
$\omega^{2}$, this order of limits gives the famous nonzero result
\begin{equation}
\lim_{q\to 0}\lim_{\omega\to 0}\sigma_{H}(\omega,q) 
= \frac{n}{B} = \frac{\nu}{2\pi}.   
\label{20}
\end{equation}
It is worth mentioning that in a normal viscous liquid as well as in a
Fermi liquid at $T=0$ we have $\mu(\omega)\sim i\omega$, which leads
to a vanishing dissipationless Hall conductivity, $\sigma_{H}(0,q)=0$.  

Using Eq.~(\ref{LLL}) one can easily show that the Hall conductivity
also determines the response of the density, $\delta n=-n\nabla {\bf
  u}$, to an external magnetic field $b=\nabla\times{\bf a}$. Indeed,
setting $a_{i}(\omega,{\bf q})=i\varepsilon_{ij}q_{j}b_{q}/q^{2}$,
and solving Eq.~(\ref{LLL}) we find
$$
\delta n_{q} = \sigma_{H}(\omega,q)b_{q}
$$
Hence a nonvanishing limit in Eq.~(\ref{20}) implies a creation of a
charge $Q=\nu \Phi/2\pi$ by an adiabatic insertion of a magnetic flux
$\Phi=b_{q=0}$. By inserting precisely one flux quantum, $\Phi=2\pi$,
(the Laughlin's gedanken experiment) we create a fractional charge
$\nu$. 

Another important quantity is the density
response function $\chi(\omega,q)$, which relates $\delta n=-n\nabla
{\bf u}$ to the external scalar potential: $\delta n=\chi(\omega,q)\varphi$.
The solution of the equation of motion, Eq.~(\ref{LLL}), with 
${\bf F}=-i{\bf q}\varphi$ yields
\begin{equation}
\chi(\omega,q) = n^{2}
\frac{\mu q^{4}}{\omega^{2}(nB + \Lambda q^{2})^{2}
- \mu(\mu + K)q^{4}}.
\label{21}
\end{equation}
Inserting Eqs.~(\ref{15}) and (\ref{16}) into Eq.~(\ref{21}) we find
that at $\omega\to 0$ and small wave vectors $\chi\sim q^{4}$, which
signifies the proper incompressibility of the FQH liquid
\cite{GirMacPla1986}.

It should be stressed out that all abovementioned fundamental low
energy properties of FQH liquids are guaranteed by the low frequency
behavior of the shear modulus, $\mu\sim\omega^{2}$. In fact, this
behavior guarantees that the low energy physics is dominated by the
Lorentz force term in the Lagrangian/equations of motion. In other
words, in the limit $\omega\to 0$ the Lagrangian Eq.~(\ref{8}) reduces
to the form
\begin{eqnarray}
L^{\rm liq} &\approx& 
- \frac{1}{2}nB{\bf u}(\partial_{t}{\bf u}\times{\bf e}_{z}) 
+ n{\bf u}\partial_{t}{\bf a} - n{\bf u}\nabla\varphi \nonumber\\
&=& - \frac{1}{2}nB\varepsilon_{ij}u_{i}\partial_{t}u_{j}
+ nu_{i}\partial_{t}a_{i} - nu_{i}\partial_{i}\varphi,
\label{CS1}
\end{eqnarray}
which is precisely a Chern-Simons Lagrangian in a temporal
gauge. Indeed, considering the standard Chern-Simons theory
\begin{equation}
L^{CS} =
\frac{\nu^{-1}}{4\pi}\varepsilon_{\alpha\beta\gamma}A_{\alpha}
\partial_{\beta}A_{\gamma} 
- \frac{1}{2\pi}\varepsilon_{\alpha\beta\gamma}A_{\alpha}
\partial_{\beta}a_{\gamma},
\label{CS}
\end{equation}
and setting $A_{i}=2\pi n\varepsilon_{ij}u_{j}$, $A_{0}=0$, and
$a_{0}=\varphi$ we recover the low frequency Lagrangian
Eq.~(\ref{CS1}). Thus our theory, defined by the Lagrangian of
Eq.~(\ref{8}), smoothly interpolates between two exactly known
limiting forms -- the low energy Chern-Simons theory, Eq.~(\ref{CS}),
and the high-frequency elasticity theory, Eq.~(\ref{3}). Below we show
that this interpolation predicts a spectrum of bulk collective excitations,
which is in excellent agreement with experimental observations. It is
worth noting that since in the low frequency limit our theory reduces
to the Chern-Simons theory, it should reproduce the gapless low energy
chiral edge excitations \cite{Wen1995}. However, the full theory
defined by Eq.~(\ref{8}) is also valid beyond the limit of
asymptotically small frequencies. Therefore it should be able to
predict a modification of the edge spectrum at higher energies.

\subsection{Spectrum of collective excitations}

To find collective modes of the system one has to set ${\bf F}=0$ in
Eq.~(\ref{LLL}), and to solve the resulting eigenvalue
problem. Alternatively, the frequencies of eigenmodes can be
determined from the poles of the density response function
$\chi(\omega,q)$, Eq.~(\ref{21}). The corresponding dispersion
equation takes the form
\begin{equation}
\omega^{2}\left[nB + \Lambda(\omega) q^{2}\right]^{2}
- \mu(\omega)\left[\mu(\omega) + K\right]q^{4} = 0.
\label{dispEq}
\end{equation}
Using the dynamic moduli $\mu(\omega)$, Eq.~(\ref{15}), and
$\Lambda(\omega)$, Eq.~(\ref{16}), and solving Eq.~(\ref{dispEq}) we
find that the density response function has only two poles. These
poles correspond to two collective modes with frequencies
$\Omega_{+}(q)$ and $\Omega_{-}(q)$. Both modes are gapped and have
the following small-$q$ dispersion
\begin{equation}
\Omega_{\pm}(q) = \Delta_{\pm} \pm \bar{\mu}_{\infty}f_{\pm} q^{2}l^{2},
\label{22}
\end{equation}
where $\bar{\mu}_{\infty}=\mu_{\infty}/n$ is the high frequency shear
modulus per particle, and $l=1/\sqrt{B}$ is the magnetic length.
Since $f_{\pm}$, Eq.~(\ref{18}), are positive, the dispersions of two
collective modes always have opposite curvatures. 

The origin of gapped collective modes can be traced back to the
dynamics of the tensor gauge field $\hat{\bm\eta}$, which is governed
by the shear Lagrangian Eq.~(\ref{7}). In fact, the gaps, $\Delta_{+}$
and $\Delta_{-}$, correspond to eigenvalues of the
operator in the left hand side of Eq.~(\ref{11}). The presence of two
modes with opposite curvatures of small-$q$ dispersion relations
agrees very well with recent experiments \cite{Hirjibehedin2005}. At
$q=0$ both gapped modes are purely quadrupole, while at any finite $q$
they have a mixed dipole-quadrupole character with an increasing dipole
component at larger $q$. Microscopically the two gaped modes are most
likely related to a mixture of small $q$ magneto-roton excitations
\cite{GirMacPla1986} and two-roton bound states
\cite{ParJai2000,GhoBas2001}.

In the limit of small wave
vectors the density response function, Eq.~(\ref{21}), takes the form
\begin{equation}
\chi(\omega,q) = \mu_{\infty}\left[
\frac{f_{-}}{\omega^{2}-\Omega_{-}^{2}(q)} + 
\frac{f_{+}}{\omega^{2}-\Omega_{+}^{2}(q)}\right]q^{4}l^{4}.
\label{23}
\end{equation}
Hence the quantities $f_{\pm}\ge 0$ determine the physical
oscillator strengths for two collective modes. Thus Eqs.~(\ref{17})
and (\ref{18}) relate all phenomenological parameters of the theory to
experimentally measurable quantities.

Strictly speaking, the Lagrangian of Eq.~(\ref{8}) defines only a long
wavelength theory. However, the results for collective modes look
quite reasonable for any value of $q$. Indeed, a straightforward
solution of Eq.~(\ref{dispEq}) shows that with increase of $q$
the function $\Omega_{+}(q)$ monotonically increases, while
$\Omega_{-}(q)$ passes through a roton-like minimum, and approaches a
constant value at $ql\to\infty$. Thus an overall behavior of the lower
branch, $\Omega_{-}(q)$, surprisingly well reproduces a physically
expected structure of the magneto-exciton spectrum, at least for the
Laughlin fractions, $\nu=1/(2k+1)$. 

The most important result of the present theory is a prediction of a
double-mode structure of gapped collective excitations. The existence
of two gapped modes is a generic fact for almost any set of
parameters, $\Delta_{1}^{2}$, $\Delta_{2}$, and $\Delta'$ (it is
somewhat more convenient to fix $\Delta_{+}$, $\Delta_{-}$, and
$\Delta'$). There is, however, a special, very narrow region in the
parameter space where the upper branch of collective excitations
disappears, and only one gapped mode with a roton minimum is left. In
this region our general construction reduces to a simplified
single-mode magnetoelasticity theory of
Ref.~\onlinecite{TokPRB2006}.  In the single-mode theory the
dynamic moduli, $\mu$ and $\Lambda$ take the
form \cite{TokPRB2006}
\begin{equation}
\mu(\omega) = \mu_{\infty}\frac{\omega^2}{\omega^2 - \Delta^2},
\quad
\Lambda(\omega) = \mu_{\infty}\frac{\Delta}{\omega^2 - \Delta^2}.
\label{single-mode}
\end{equation}
Using Eqs.~(\ref{15})-(\ref{18}) we find that expressions of
Eq.~(\ref{single-mode}) are recovered in two limiting cases: (1)
$\Delta_{+}\to\infty$, $\Delta_{-}=\Delta$, and  $\Delta'$ is an
arbitrary finite constant; (2)
$\Delta_{+}=\Delta_{-}=\Delta'=\Delta$. 
In this single-mode region of the parameter space an effective shear
Lagrangian, Eq.~(\ref{7}), reduces to the following simple form
$$
L^{\rm liq}_{\rm S} =       
- \frac{1}{4}\mu_{\infty}{\rm Tr}(\hat{\bf S} - \hat{\bm\eta})^{2}
+ \frac{\mu_{\infty}}{4\Delta}
{\rm Tr}\hat{\bm\eta}(\partial_{t}\hat{\bm\eta}\times{\bf e}_{z}).
$$
All the rest of the three-dimensional parameter space,
$(\Delta_{+},\Delta_{-},\Delta')$, corresponds to a theory with
two gapped collective modes.
It has been shown in Ref.~\onlinecite{TokPRB2006} that the
single-mode magnetoelasticity can be derived from the fermionic
Chern-Simons theory
\cite{LopFra1991,LopFra1993,SimHal1993,CompositeFermions} within the
random phase approximation (RPA). In the present section we have
demonstrated that the symmetry allows for a more general construction,
which leads to a double-mode excitation spectrum. Microscopically this
general FQH hydrodynamics should most likely emerge from post-RPA
vertex corrections.

\section{Hydrodynamics of quantum Hall crystal and liquid crystal
  phases} 

In Sec.~II we derived a hydrodynamic theory of incompressible quantum
Hall liquids. In the present section this approach is generalized to
non-liquid collective states. The hydrodynamics of FQH liquids was
based on two principal assumptions, (i) and (ii). Note that at the
long wavelength, i.~e. for small $q$, in 2D systems the rotational
symmetry is indistinguishable from the hexagonal one (with an accuracy
up to $q^{2}$). Hence in the long wavelength limit the assumption (i)
also covers all collective states of hexagonal symmetry. Let us keep
the assumption (i), but relax (ii) by a successive reduction of the
diffeomorphism symmetry defined by Eq.~(\ref{1}).

\subsection{Hexatic liquid crystal} 

On the first stage we reduce the general {\em local} symmetry of
Eq.~(\ref{1}) to the invariance under {\em global} shear
deformations. The later correspond to transformations Eq.~(\ref{1})
with $\varphi({\bf r})$ being a second order polynomial of
$r_{i}$. Such transformations produce a constant in space correction
$\delta\hat{\bm S}$, Eq.~(\ref{6}), to the shear strain tensor. The
rotational symmetry combined with the global invariance under the
transformations Eq.~(\ref{1}) allow only one term to be added to the
shear Lagrangian Eq.~(\ref{7}),
\begin{equation}
\delta L_{\rm S}^{\rm hex} = - \frac{\kappa}{4}{\rm Tr}
(\partial_{i}\hat{\bm\eta})(\partial_{i}\hat{\bm\eta}),
\label{24}
\end{equation}
where $\kappa$ is a positive constant.  The full Lagrangian of a
collective state that is invariant with respect to global shear
deformations takes the form
\begin{equation}
L^{\rm hex} = L_{0} + L^{\rm liq}_{\rm S}+\delta L_{\rm S}^{\rm hex} 
+ {\bf u}\partial_{t}{\bf a} - {\bf u}\nabla\varphi.
\label{25}
\end{equation}

Apparently an additional term, $\delta L_{\rm S}^{\rm hex}$, in the
Lagrangian influences only the equation of motion for the gauge field
$\hat{\bm\eta}$. Hence the equation for the
displacement vector, Eq.~(\ref{9}), as well as the relation of
Eq.(\ref{10}) remain unchanged. Only the equation of motion for
the shear stress tensor $\hat{\bm\pi}$ gets modified and acquires
gradient corrections 
\begin{eqnarray}\nonumber
\partial_{t}^{2}\hat{\bm\pi} &+& \Delta_{1}^{2}\left[
1- \frac{\kappa}{\mu_{1}}\nabla^{2}\right]\hat{\bm\pi} 
- \Delta_{2}\partial_{t}\hat{\bm\pi}\times{\bf e}_{z} = 
-\mu_{\infty}\partial_{t}^{2}\hat{\bf S}\\ 
&+& \Delta_{1}^{2}\kappa\nabla^{2}\hat{\bf S} -
\mu_{1}(\Delta_{2} + 2\Delta')\partial_{t}\hat{\bf S}\times{\bf e}_{z}
\label{piHex}
\end{eqnarray}
[we note that Eq.~(\ref{13}) still holds, which follows from the
universality of the $\omega\to\infty$
asymptotics]. 

The rotational, and the time-reversal symmetries require
\cite{TokPRB2006} the solution of Eq.~(\ref{piHex}) to be of the
general form given by Eq.~(\ref{14}). Hence the strong field ($m\to
0$) expressions for the Hall conductivity and for the density response
function, Eqs.~(\ref{19}) and (\ref{21}), as well as the dispersion
equation, Eq.~(\ref{dispEq}), do not change. The analytic structure
of the elastic moduli $\mu$ and $\Lambda$ is, however,
different. Due to the gradient terms in Eq.~(\ref{piHex}) the dynamics
moduli become functions of both the frequency $\omega$ and the wave vector
$q$.

The most important effect of the gradient term, $\delta L_{\rm S}^{\rm
hex}$, in the Lagrangian of Eq.~(\ref{25}) is a nonzero value of the
shear modulus $\mu(\omega,q)$ at $\omega=0$ and finite $q$. Indeed,
neglecting all time derivatives in Eq.~(\ref{piHex}) we get the
following static equation of state
$$
\hat{\bm\pi} = -\frac{\kappa q^{2}}{1+\kappa q^{2}/\mu_{1}}\hat{\bf S} 
= - \mu(0,q)\hat{\bf S}.
$$
It the long wavelength limit the static shear modulus $\mu_{0}(q)$
takes the form 
\begin{equation}
\mu_{0}(q)=\mu(0,q) \approx \kappa q^{2}
\label{26}
\end{equation}
(by construction the high-frequency limit of $\mu(\omega,q)$ is fixed
to the value $\mu_{\infty}$).  The static shear modulus Eq.~(\ref{26})
vanishes in the limit $q\to 0$. Therefore the state described by the
Lagrangian Eq.~(\ref{25}) can not be a solid. To identify the nature
of this state we calculate the equal-time correlation function of the
orientational order parameter $\Psi({\bf r}) = e^{i\nabla\times{\bf
u}({\bf r})}$.\cite{NelHal1979} At finite temperature $T$ the behavior
of the correlation function $\langle\Psi^{*}({\bf r})\Psi(0)\rangle$
at $r\to\infty$ is determined by the static part of the Lagrangian
$L^{\rm hex}$. Using an explicit form of Eq.~(\ref{25}) (without
external fields), and performing the standard calculations we arrive
at the following result
\begin{equation}
\langle\Psi^{*}({\bf r})\Psi(0)\rangle \sim
 r^{-\frac{n^{2}T}{2\pi\kappa}}.
\label{27}
\end{equation}
An algebraic decay of the correlator Eq.~(\ref{27}) is a clear signature
of a hexatic liquid crystal \cite{NelHal1979}.

A nonzero static shear modulus $\mu_{0}$ of hexatic state dramatically
changes the low energy physics. Using the general formulas for the
response functions, Eqs.~(\ref{19}) and (\ref{21}), and the limiting
form of the shear modulus, Eq.~(\ref{26}), we get
\begin{equation}
\lim_{q\to 0}\lim_{\omega\to 0}\sigma_{H}(\omega,q) = 0, \quad
\lim_{q\to 0}\lim_{\omega\to 0}\chi(\omega,q) = - \frac{n^{2}}{K}.    
\label{27a}
\end{equation}
Thus violation of the local symmetry of Eq.~(\ref{1}) destroys the
dissipationless quantum Hall effect, and recovers the standard form of
the compressibility sum rule \cite{PinNoz1}.

The low energy behavior of the response functions, Eq.~(\ref{27a}), is
closely related to a dramatic modification of the structure of
collective modes. In addition to two gapped modes, which remain
practically unchanged in comparison with Eq.~(\ref{22}), we find one
more nontrivial gapless solution to the dispersion equation,
Eq.~(\ref{dispEq}). The frequency of the gapless mode at small $q$ is
readily obtained by inserting the static shear modulus $\mu_{0}$,
Eq.~(\ref{26}), into Eq.~(\ref{dispEq}):
$$
\Omega_{0} = n^{-1}\sqrt{\mu_{0}(\mu_{0}+K)}q^{2}l^{2}
$$
In a system with Coulomb interaction the small-$q$ behavior of the
bulk modulus is dominated by the Hartree contribution, $K\approx 2\pi
n^{2}/q$.  \cite{note2} In this case the dispersion of the gapless mode in
the hexatic phase takes the form
$$
\Omega_{0}^{\rm hex}(q) = B^{-1}\sqrt{2\pi\kappa}q^{5/2}
$$
(for a short range interaction we get $\Omega_{0}^{\rm hex}\sim
q^{3}$). Interestingly, the above dispersion law is similar to that
predicted for a Goldstone mode in a quantum Hall nematic phase
\cite{RadDor2002}. 

Recently a number of microscopic trial wave functions for different
quantum Hall
liquid crystal states have been proposed \cite{CifLapWex2004}. The
single mode approximation (SMA) \cite{GirMacPla1986} applied to the
hexatic wave function yields a gapped collective mode with a roton
minimum \cite{LapWex2006}. This behavior excellently
correlates with our results for the lower gapped mode, $\Omega_{-}(q)$
-- the dispersion of this mode has practically the same form both for
hexatic and for a liquid states. The microscopic SMA fails to produce
a gapless mode that should be present in liquid crystal phases
\cite{LapWex2006}. In contrast, our approach simultaneously
predicts both the gapped modes and the dispersion of the Goldstone
mode for the quantum Hall hexatic liquid crystal.

\subsection{Crystalline states}

Finally we completely destroy any kind of invariance under shear
deformations. The most radical way to do that is to drop out the gauge
field $\hat{\bm\eta}$ by setting $\Delta_{1},\Delta_{2}\to 0$ in
Eq.~(\ref{7}). The resulting Lagrangian is simply $L^{\infty}$,
Eq.~(\ref{3}). It describes the long wavelength dynamics of a hexagonal 2D
solid with a frequency independent shear modulus
$\mu=\mu_{\infty}$. Apparently this state corresponds to a Wigner
crystal. In the limit of a strong magnetic field ($m\to 0$) we get
only one collective mode that is a well known \cite{BonMar1977}
magneto-phonon with the dispersion
$$
\Omega_{0}^{\rm W} = B^{-1}\sqrt{2\pi\mu_{\infty}}q^{3/2}.
$$ 

A more delicate way to violate the symmetry of Eq.~(\ref{1}) is to add
to Eq.~(\ref{25}) the last invariant, ${\rm Tr}(\hat{\bm\eta})^{2}$,
that is allowed by the rotational and the time-reversal
symmetries. The resulting total Lagrangian takes the form
\begin{equation}
L^{\rm bub} = L^{\rm hex}-\frac{1}{4}\mu_{2}{\rm Tr}(\hat{\bm\eta})^{2},
\label{Lbub}
\end{equation}
where $L^{\rm hex}$ is defined after Eq.~(\ref{25}), and $\mu_{2}$ is
an additional phenomenological constant. The second term in
Eq.~(\ref{Lbub}) leads to the following simple modification of the
equation of motion for the shear stress tensor $\hat{\bm\pi}$: in
Eq.~(\ref{piHex}) the term $-\kappa\nabla^{2}$ is replaced by the
combination $-\kappa\nabla^{2} + \mu_{2}$. Noting this fact we easily
realize that the Lagrangian of Eq.~(\ref{Lbub}) describes a crystal
state with a constant static shear modulus of the form
$$
\mu_{0}= \mu(0,0) = \frac{\mu_{1}\mu_{2}}{\mu_{1} + \mu_{2}},
$$ 
where $\mu_{1}$ is related to $\mu_{\infty}$ by Eq.~(\ref{13}). 

A nonzero value of $\mu_{0}$ leads to a gapless magneto-phonon mode
with a small-$q$ dispersion similar to that for a classical Wigner crystal 
$$
\Omega_{0}^{\rm bub} = B^{-1}\sqrt{2\pi\mu_{0}}q^{3/2}.
$$
In addition, the dynamics of the gauge field $\hat{\bm\eta}$, which is
encoded in the equation of motion for the stress tensor
$\hat{\bm\pi}$, still produce two FQH liquid-like gapped modes. The
corresponding $q=0$ gaps, $\Delta_{+}^{\rm bub}$ and $\Delta_{-}^{\rm
  bub}$, are related to parameters of the Lagrangian as follows
$$
\Delta_{+}^{\rm bub}\Delta_{-}^{\rm bub} = 
\Delta_{1}^{2}\left(1+\frac{\mu_{2}}{\mu_{1}}\right), \quad
\Delta_{+}^{\rm bub} - \Delta_{-}^{\rm bub} = \Delta_{2}
$$
It is natural to interpret this intermediate state as a triangular
lattice of liquid ``bubbles''. The described structure of collective
modes is indeed in a qualitative agreement with the Hartree-Fock
excitation spectrum of a quantum Hall bubble phase \cite{Cote2003}.

\section{Conclusion}

In conclusion we proposed a unified hydrodynamical framework for the
description of collective quantum Hall states. In particular, the
present theory covers incompressible FQH liquids, hexatic liquid
crystals, and two different hexagonal crystal phases that can be
identified as a Wigner crystal, and a bubble solid. We predicted the
dispersion of the Goldstone mode for quantum Hall hexatics,
$\Omega_{0}^{\rm hex}\sim q^{5/2}$, and the presence of two gapped
modes for incompressible quantum Hall liquids. The later result
naturally explains recent experimental observations of the long
wavelength dispersion of collective modes in the $\nu=1/3$ FQH state
\cite{Hirjibehedin2005}.

In short, the hydrodynamics of FQH liquids is based on the following
observations. On the one hand, the high-frequency response of any
system is universally elastic. On the other hand, a liquid state is
insensitive to static volume preserving deformations. The former
requirement fixes the high-frequency form of the theory, while the
later one implies a particular local symmetry which can be implemented
using a proper gauge field. Since our approach relies only on
symmetry, we believe that the predicted two gapped modes should be a
universal property of a 2D collective magnetized liquid.

\section*{Acknowledgment}
I am grateful to G. Vignale for the
encouragement and for numerous illuminating discussions.


\begin{thebibliography}{30}
\bibitem[{not({\natexlab{c}})}]{note2}
\bibinfo{note}{To shorten formulas the Hartree term is omitted in
  Eq.~(\ref{4}). When necessary, the Hartree contribution is easily restored by
  the replacement $K\to K + n^2V(q)$, where $V(q)$ is the interaction
  potential.}
\end{thebibliography}

\end{document}